\documentclass[conference]{IEEEtran}
\ifCLASSINFOpdf
\else
\fi
\usepackage{amsmath}
\usepackage{graphicx}

\begin{document}
\title{Structured Optical Receivers for Efficient Deep-Space Communication}

\author{
\IEEEauthorblockN{Konrad Banaszek and Micha\l{} Jachura}
\IEEEauthorblockA{Centre of New Technologies\\
University of Warsaw\\
PL-02-097 Warszawa, Poland\\
E-mail: k.banaszek@cent.uw.edu.pl}}


%Telephone: (800) 555--1212, Fax: (888) 555--1212}


\maketitle

\begin{abstract}
We discuss conceptual designs for structured optical receivers that can alleviate the requirement for high peak-to-average power ratio in photon-starved optical communication. The basic idea is to transmit sequences of suitably modulated coherent light pulses whose energy can be concentrated in a single temporal bin on the receiver side through optical interference. Two examples of scalable architectures for structured receivers are presented. The first one, based on active polarization switching, maps Hadamard codewords composed from the binary phase shift keying (BPSK) constellation onto the standard pulse position modulation (PPM) format. The second receiver, using solely passive optical elements, converts phase-polarization patterns of coherent light pulses into a single pulse preserving a synchronized time of arrival. Such a conversion enables implementation of a communication protocol equivalent to the PPM scheme but with distributed optical power provided that the intersymbol guard-time exceeds the pattern length.
\end{abstract}


% \fi
\IEEEpeerreviewmaketitle

\section{Introduction}
The pulse position modulation (PPM) format, routinely used in space optical communication, requires an unbalanced distribution of instantaneous light intensity \cite{Waseda2011,Moision2014}. This results in stringent demands on the peak-to-average power of laser light source used in the transmitter. A common design for the transmitter relies on carving the PPM signal out of a continuous wave laser beam using an electro-optic amplitude modulator and subsequently amplifying it with the help of e.g.\ an erbium-doped fiber amplifier. This architecture, known as the master-oscillator power amplifier (MOPA) \cite{Caplan}, offers multigigahertz bandwidth limited by the speed of electro-optic modulators and photodetectors. The peak power of several watts offered by the MOPA architecture is sufficient for relatively short communication links such as Earth-LEO, Earth-GEO or Earth-Moon \cite{LunarI,LunarII}.

In deep-space communication scenarios optical peak power in the kilowatt range is indispensable and Q-switched laser transmitters are regularly used \cite{Hemati}. Although they are able to deliver the desired peak power, the pulse repetition rate attainable in the Q-switching technique is limited to approx.\ $200$~kHz, which compared to several GHz provided by MOPA significantly reduces practical transmission rates. Due to several detrimental factors such as the non-unit transfer efficiency of pump power into the active medium and the limited optical-to-optical conversion efficiency of the active medium, the wall-plug efficiency of Q-switched laser transmitters does not usually exceed $15\%$ and is in general much lower than that attainable in the MOPA design. Furthermore, peak power levels can exceed the damage threshold of optical components in the transmitter setup and may lead to heat dissipation issues.

The requirements on the transmitter laser source can be significantly altered through the use of recently proposed structured optical receivers. The basic idea is to spread the signal optical energy over multiple bins and to perform joint detection on sequences of incoming elementary symbols \cite{GuhaI}. The advantage offered by this strategy is known in quantum information science as the {\em superadditivity} of accessible information. A class of joint-detection schemes proposed by Guha \cite{GuhaII} uses the well-known binary phase shift keying (BPSK) constellation to construct so-called Hadamard codewords. On the receiver side, such codewords can be converted using a linear optical circuit into the standard PPM format suitable for direct detection. Consequently, the spectral efficiency of the PPM scheme \cite{Wang2014,Jarzyna2015} can in principle be attained with uniform distribution of the signal optical power and phase modulation only in the transmitter setup. In this paper, we describe a conceptual design for a scalable receiver architecture that realizes conversion from Hadamard codewords into the PPM format using active polarization switching. We also present a second type of a structured optical receiver, based on passive optical elements, that concentrates the entire optical energy of a phase-polarization pattern of incoming light pulses in a single time bin. With a sufficiently long intersymbol guard-time, such a receiver offers the performance of the PPM scheme with optical energy distributed over multiple time bins.

\section{Active receiver for Hadamard sequences}

\begin{figure*}
\includegraphics[width=0.975\textwidth]{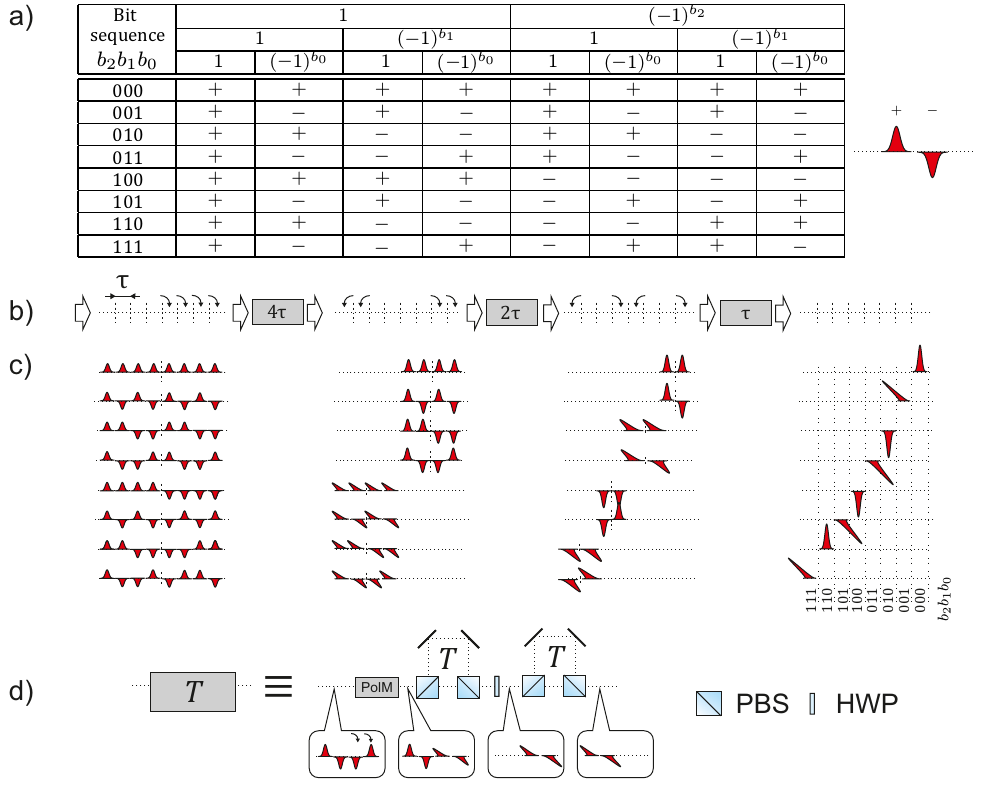}\centering\caption{
Active optical receiver converting Hadamard codewords of length $2^m$ composed from the $\pm$ BPSK constellation into the $2^m$-ary PPM format, shown for $m=3$. (a) Construction of Hadamard codewords through a hierarchy of sign relations. Each entry is a product of $\pm 1$ factors above it. (b) Scalable architecture for a structured receiver consisting of active modules using delays $2^{m-1}\tau$, $\ldots$, $2\tau$, $\tau$, where $\tau$ is the time bin separation. Curved arrows indicate polarization switching implemented by the polarization modulator in each module. (c) Transformation of individual Hadamard codewords at the consecutive stages of the receiver. (d) Setup of an individual module for a delay $T$. PolM, polarization modulator; PBS, polarizing beam splitter; HWP, half-wave plate. The callouts illustrate the transformation of an exemplary Hadamard sequence $+--+$ for $T=2\tau$.
 \label{fig:ActiveLoopy}}
\end{figure*}

The superadditive communication scheme proposed in \cite{GuhaII} uses codewords composed from the BPSK constellation whose two elements will be denoted symbolically as $+$ and $-$. The codewords are chosen as rows of a Hadamard matrix. Hadamard matrices are orthogonal symmetric matrices with entries $\pm 1$ that exist for dimensions equal to integer powers of $2$ and specific other natural numbers. For a Hadamard matrix of dimension $2^m$, each row can be identified bijectively with a bit string $b_{m-1} \ldots b_{1} b_{0}$ of length $m$, as illustrated with Fig.~\ref{fig:ActiveLoopy}(a) for $m=3$. The bits define a hierarchy of $\pm$ sign relations between binary sections within a row. An individual entry in a row is given by a product of all $\pm 1$ factors above it as shown in Fig.~\ref{fig:ActiveLoopy}(a).

The chosen codeword is used to modulate a sequence of $2^m$ coherent light pulses. On the receiver side, the pulses are interfered using a linear optical circuit such that for each codeword the entire optical energy is concentrated in a different output port of the circuit. In the exemplary implementation presented in the original proposal \cite{GuhaII}, pulses are fed into distinct spatial input ports of the circuit. If the received signal has the form of a pulse sequence arriving along one spatial path in consecutive time bins, using such a circuit would require prior rerouting and delaying individual pulses. Alternatively, the linear optical transformation could be in principle realized  in a multimode quantum memory system \cite{Klimek}, but this technology is still in its infancy.

\begin{figure*}%%[th!]
\includegraphics[width=0.975\textwidth]{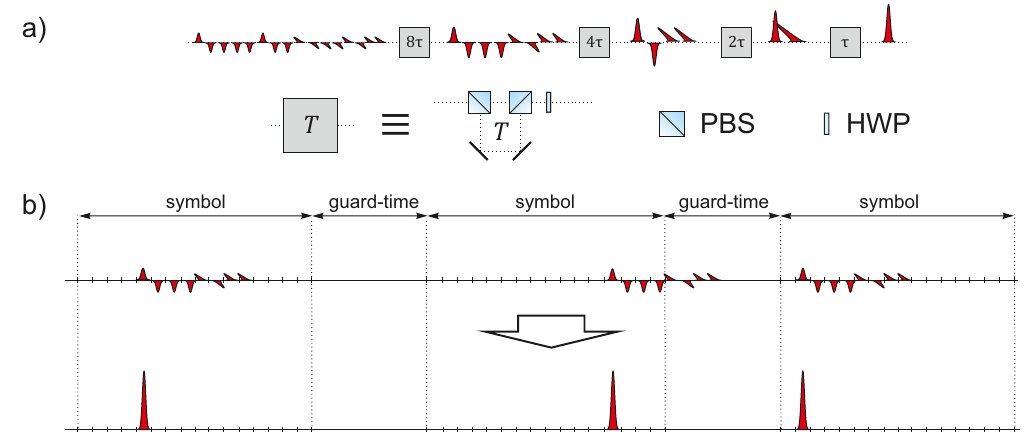}\centering\caption{Passive optical receiver producing single pulses from phase-polarization patterns. (a) The scalable architecture for patterns spanning $2^m$ bins shown for $m=4$. Each module represented by a square box and labelled with a delay $T$ consists of a polarization-selective delay line followed by a half-wave plate. (b) The phase-polarization pattern for $m=3$ used in a 16-ary PPM format. The peak power requirement is reduced eightfold. \label{fig:PassiveLoopy}}
\end{figure*}

A scalable architecture based on active polarization switching that maps temporal Hadamard codewords directly onto the PPM format is presented in Fig.~\ref{fig:ActiveLoopy}(b) for $m=3$. The architecture consists of a series of modules represented by rectangular boxes.
Fig.~\ref{fig:ActiveLoopy}(c) shows the transformation of individual Hadamard codewords by the modules.
The internal setup of one module is depicted in Fig.~\ref{fig:ActiveLoopy}(d).  The first element of the module is a polarization modulator PolM which switches portions of the input signal between horizontal and vertical polarizations according to curved arrows shown in Fig.~\ref{fig:ActiveLoopy}(b) before each module. The switching pattern is independent of the input codeword. Next, the horizontal component of the signal is delayed with respect to the vertical one by time $T$ specified for each module. This step can be realized using two polarizing beam splitters PBS and a delay line. As a result, horizontal and vertical components of the signal overlap in the same time bins. Because of fixed phase relations between these components which follow from the hierarchical construction of Hadamard codewords presented in Fig.~\ref{fig:ActiveLoopy}(a), the signal in each bin will have one of two diagonal polarizations oriented at $+45^\circ$ or $-45^\circ$. The diagonal polarizations are brought to the rectilinear basis of the horizontal and the vertical polarizations with the help of a half-wave plate HWP. In the last step, the horizontal polarization is delayed with respect to the vertical one in another delay line introducing delay $T$ for the horizontal polarization. This reduces the length of the sequence by a factor of two and doubles the intensity of each output light pulse.

For Hadamard sequences of length $2^{m}$ and the temporal spacing between consecutive bins equal to $\tau$, the delay of the the first module is $2^{m-1}\tau$. The delay of each subsequent module is halved until the single-bin delay $\tau$ is reached. Thus $m$ modules are needed altogether, which makes the architecture scaling logarithmically in the codeword length. It is easy to see that at the output of the last module the energy of the entire Hadamard sequence is concentrated in a single time bin whose position is given by the sum $\sum_{k=0}^{m-1}b_{k}2^{k}$, i.e.\ the natural number specified in the binary representation by the bit sequence $b_{m-1} \ldots b_1 b_0$. Thus Hadamard codewords are converted into the PPM symbols that can be identified using direct detection from the timing of photocount events. Instances when no detection event is recorded over the entire Hadamard sequence are interpreted as erasures in a full analogy with the standard PPM communication scheme.
Note that one can also devise a tree-like architecture built from modules that comprise a polarization modulator and a single delay stage. At the output of each module the two diagonal polarizations are separated, rotated to the vertical polarization, and fed into a pair of modules with half the delay. Such an architecture converts Hadamard sequences into the spatial PPM format, but it would require $2^m-1$ modules with active polarization switching.

\section{Passive receiver for phase-polarization patterns}

An altenative option to reduce the requirement for the high peak-to-average power ratio on the transmitter side is to generate phase-polarization pulse patterns that span multiple time bins. We will demonstrate that with a suitable choice of phase and polarization relations between individual pulses in a pattern, such patterns can be converted using a structured receiver build only from passive optical elements into one pulse occupying a single time bin.

The above idea is shown in Fig.~\ref{fig:PassiveLoopy}(a) for a pattern spanning $2^m$ bins with $m=4$.
The conversion takes place in a sequence of modules characterized by delays $2^{m-1}\tau$, $\ldots$, $2\tau$, $\tau$, where $\tau$ is the time bin spacing. First, each module delays the horizontally polarized components of the signal with respect to the vertical ones in order to attain a two-fold contraction of the pattern length. The coherently superposed components, emerging from the delay stage at $\pm 45^\circ$ polarizations, are rotated by a half-wave plate to the rectilinear basis before entering the next stage. After all $m$ stages, the sequence of modules produces a pulse in a single time bin that carries the optical energy of the entire pattern. Its temporal location can be detected by photon counting.
Information is encoded in symbols with different pattern arrival times analogously to the standard PPM format as shown in Fig.~\ref{fig:PassiveLoopy}(b) for $m=3$ and the PPM order equal to $16$. In order to avoid an overlap between patterns, the guard-time between consecutive symbols needs to be longer than the pattern length.

An elementary way to determine polarizations and phases of individual pulses in the pattern is to analyze the action of the presented circuit in reverse, when the output port is hypothetically injected with a single pulse. Note that this reverse mode has been previously used for time-multiplexed photon number resolved detection using Geiger-mode operated avalanche photodiodes, with horizontal and vertical  polarizations corresponding to a pair of distinct optical paths in single-mode fibers \cite{Achilles2003,Fitch2003}. The pattern can be generated from a laser beam polarized at $45^\circ$ using two crossed phase modulators modulating respectively the horizontal and the vertical components. The beam needs to be blocked outside the timespan of the pattern. If the symbol frame is of the same duration as the intersymbol guard-time, this results in the effective $50\%$ usage of the optical power.
Including a second pattern corresponding to the orthogonal polarization at the circuit output doubles the number of available PPM symbols. In order to allow rotational freedom between the transmitter and the receiver, polarizations labelled as horizontal and vertical should be converted for transmission into circular ones using a quarter wave plate.
\enlargethispage{2.28in}

\section{Discussion}

The demand for high peak-to-average power on the transmitter side in photon-starved optical communication can be in principle alleviated through the use of structured optical receivers. In this approach, the transmitter emits sequences of mutually coherent light pulses modulated in phase and optionally also in polarization. The receiver combines pulses using optical interference to concentrate the energy of the entire sequence into a single time bin. The information is encoded in the position of the produced single pulse, which provides efficiency comparable to that of the standard PPM format. While the complication of the transmitter setup can be viewed as relatively minor, the construction and the operation of a structured optical receiver poses a \enlargethispage{-3.5in} number of technical challenges. The receiver needs to accommodate spatial and temporal mode distortions that may occur in the course of propagation through the optical channel, hence its construction should have inherent multimode capability. Significant progress has been made towards tolerance of distortions in interferometric receivers in the context of free-space quantum key distribution \cite{Waterloo2015}. All delays need to be stabilized to a fraction of the wavelength to ensure correspondingly constructive or destructive interference in individual time bins. In order to ensure high visibility of the optical interference, distortions introduced by the optical channel need to fluctuate over a time scale that exceeds substantially the duration of entire codewords or phase-polarization patterns. Also, one can expect that optical interference will be perturbed at the edges of individual time bins due to the finite switching time of electro-optic modulators employed in the transmitter and the receiver. A possible solution would be to use as the light source a pulsed laser with the repetition rate corresponding to the time bin separation. This would concentrate the optical power at the center of the bins, where modulators are in a settled state.

\section*{Acknowledgment}
The authors would like to thank F.~E.~Becerra, C.~Marquardt, M.~Shaw, and W.~Wasilewski for insightful discussions.
This work is part of the project ``Quantum Optical Communication Systems'' carried out within the TEAM
programme of the Foundation for Polish Science co-financed by the European Union under the European
Regional Development Fund.



%\bibliographystyle{IEEEtran}

\end{document}